\documentstyle[12pt,epsfig]{article}
\newlength{\dinwidth}
\newlength{\dinmargin}
\newcommand\ben{\begin{equation}}
\newcommand\een{\end{equation}}
\newcommand\bea{\begin{eqnarray}}
\newcommand\eea{\end{eqnarray}}

\newcommand\vx{{\vec x}}
\newcommand\bg{{\bar g}}

\newcommand\Tr{{\rm Tr}}
\newcommand\nn{\nonumber}

\newcommand\del{\partial}
\newcommand\bA{\bar A}
\newcommand\bF{\bar F}

\begin{document}
\thispagestyle{empty}
\addtocounter{page}{-1}
\vskip-0.35cm
\begin{flushright}
UK/07-04 \\
\end{flushright}
\vspace*{0.2cm}
\centerline{\Large \bf Holography and Cosmological Singularities
\footnote{Based on talks given at CTP Symposium, Cairo, Egypt (March, 2007) and
Sowers Theoretical Physics Workshop, Blacksurg, Virginia, USA. (May, 2007)}}
\vspace*{1.0cm} 
\centerline{\bf Sumit R. Das}
\vspace*{0.7cm}
\centerline{\it Department of Physics and Astronomy,}
\vspace*{0.2cm}
\centerline{\it University of Kentucky, Lexington, KY 40506 \rm USA} 
\vspace*{1cm}
\centerline{\tt das@pa.uky.edu}

\vspace*{0.8cm}
\centerline{\bf Abstract}
\vspace*{0.3cm}
\vspace*{0.5cm}
Certain null singularities in ten dimensional supergravity have
natural holographic duals in terms of Matrix Theory and 
generalizations of the AdS/CFT correspondence. In many situations
the holographic duals appear to be well defined in regions where
the supergravity develops singularities. We describe some recent 
progress in this area.
\baselineskip=18pt
\newpage

\section{Introduction}

This talk is based on work done in collaboration with Jeremy Michelson, K. Narayan and Sandip Trivedi \cite{Das:2005vd,
Das:2006dr, Das:2006pw, Das:2006dz}.

Space-like and null singularities pose a peculiar
puzzle. At these singularities, "time" begins or ends - and it is not
clear what is the meaning of this. Classic examples of such
singularities are those which appear in the interior of neutral black
holes and those which appear in cosmology.

It has been always suspected that near singularities usual notions of
space and time break down and a consistent quantization of gravity
would provide a more abstract structure which replaces
space-time. However we do not know as yet what this abstract structure could
be in general.  In some situations, String Theory has provided concrete ideas
about the nature of this structure. These are situations where
gravitational physics has a tractable holographic description 
\cite{'tHooft:1993gx} in terms of a
non-gravitational theory in lower number of space-time dimensions. In
view of the spectacular success of the holographic principle in black
hole physics, it is natural to explore whether this can be used to
understand conceptual issues posed by singularities.

In String Theory, holography is a special case of a more general
duality between open and closed strings. This duality implies that the
dynamics of open strings contains the dynamics of closed
strings. Since closed strings contain gravity, space-time questions
can be posed in the open string theory which does not contain gravity
and therefore conceptually easier. Under special circumstances, the
open string theory can be truncated to its low energy limit - which is
a gauge theory on a {\em fixed} background.  In these situations,
open-closed duality becomes particularly useful.  The simplest example
is non-critical closed string theory in two space-time
dimensions. Here the holographic theory is gauged Matrix Quantum
Mechanics \cite{twod}. 
The second class of examples involve string theory or M
theory defined on spacetimes with a compact null direction. Then a
sector of the theory with some specified momentum in this null
direction is dual to a $d+1$ dimensional gauge theory, where $d$
depends on the number of additional (spacelike) compact directions.
Using standard terminology we will call them Matrix theories
\cite{Banks:1996vh}-\cite{Dijkgraaf:1997vv}. Finally,
the celebrated AdS/CFT correspondence \cite{Maldacena:1997re}
relates closed string theory in
asymptotically anti-de-Sitter spacetimes to gauge theories living on
their boundaries. In all these cases, the dynamical "bulk" spacetime
(on which the closed string theory lives) is an approximation which
holds in a specific regime of the gauge theory. In this regime, the
closed string theory reduces to supergravity. Generically, there is no
space-time interpretation, though the gauge theory may make perfect
sense. This fact opens up the possibility that in regions where the
bulk gravity description is singular, one may have a well defined
gauge theory description and one has an answer to the question : {\em
What replaces space-time ?}

Treating time dependent backgrounds in string theory, particularly
those with singularities, has been notoriously difficult. However, some
modest progress has been made recently in both worldsheet formulations
as well as holographic formulations of all the three types mentioned
above. The key idea in these various types of holography are similar. One looks for toy models where the space-time background on
which the closed string theory is defined is singular, but the
holographic gauge theory description does not appear to be
problematic. Thus, the gauge theory provides the correct description of
the region which would appear singular if the gravity interpretation
is extrapolated beyond its regime of validity.

In the following, we will discuss recent attempts to understand
cosmological singularities using Matrix theories as well as AdS/CFT
correspondence \footnote{For discussions of cosmological singularities in the Matrix Model description of two dimensional string theory,
see \cite{twodsingular}}.

\section{Matrix Big Bangs}

In \cite{Craps:2005wd}, Craps et. al. 
considered Type IIA string theory with 
string coupling $g_s$ and string length $l_s$, living on a
flat string frame metric with a compact null direction $x^-$ with radius $R$
\begin{equation}
ds^2 = 2 dx^+dx^- + d\vx \cdot d\vx,
\label{flatmetric}
\end{equation}
and a dilaton linearly proportional
to the other null direction $x^+$
\begin{equation}
\Phi = -Q x^+,
\label{nulldilaton}
\end{equation}
As a supergravity solution, this background
preserves half of the supersymmetries which satisfy $\Gamma^+ \epsilon = 0$.
For $Q > 0$, the effective string coupling 
$\bg_s = g_s e^{-Qx^+}$
is small for $x^+ \rightarrow \infty$ and one should have a
perturbative spectrum, while
for $x^+ \rightarrow -\infty$ the string theory becomes strongly coupled
and the corresponding Einstein metric has a null big bang like singularity \footnote{For $Q < 0$ we have a time-reversed situation where the big bang is
replaced by the 
big crunch. In this paper we will exclusively deal with $Q > 0$.}.

For $Q=0$, DLCQ string theory in this background with a momentum 
\ben
p_- = \frac{J}{R}
\een
is dual to a $1+1$ dimensional $U(J)$ gauge theory - usually
called Matrix String Theory \cite{Motl:1997th}-\cite{Dijkgraaf:1997vv} -  living on a circle of radius
${\tilde R}$ given by
\begin{equation}
{\tilde R} = \frac{l_s^2}{R};
\end{equation}
and a Yang-Mills coupling given by
\begin{equation}
g_{YM} = \frac{R}{g_s l_s^2}.
\end{equation}
The bosonic part of the
gauge theory action is
\ben
S = \int d\tau \int_0^{2\pi{\tilde R}}d\sigma {\rm Tr}  \{
\frac{1}{2 g_{YM}^2} F_{\tau \sigma}^2
+ \frac{1}{2} (D_\tau X^i)^2 
- \frac{1}{2} (D_\sigma X^i)^2
+ \frac{g_{YM}^2}{4} [ {X^i} , {X^j} ]^2 \} 
\label{mone}
\een
where $X^i, i = 1\cdots 8$ are adjoint scalars.
The above relations show that when the original string coupling
is small, $g_s \ll 1$, the Yang-Mills coupling is large and the
theory flows to the IR. The potential term then restricts the
$X^i$ to belong to a cartan subalgebra and may be therefore 
chosen to be diagonal 
\ben
X^i = {\rm diag}( X^i_1, X^i_2, \cdots X^i_J)
\een
in a suitable gauge. The gauge field 
decouples, and one is left with $8J$ scalar fields $X^i_n$ in $1+1$
dimensions. The boundary conditions of these fields are 
labelled by the conjugacy classes of the group. For example, the
maximally twisted sector has
\ben
X^i_n (\sigma + 2\pi) = X^i_{n+1}(\sigma)
\een
where $X^i_{J+1} \equiv X^i_1$. In this sector we therefore have 
8 scalars on a circle of size $2\pi l_s^2 \frac{J}{R}$ and the
action then reduces to the worldsheet action of a {\em single}
string in a light cone gauge. As is appropriate in the light cone
gauge, the spatial extent of the worldsheet is proportional to 
the longitudinal momentum $p_-$. In a similar way one has 
boundary conditions with cycles of smaller length - these sectors
represent multiple strings. Effects of finite $g_{YM}{\tilde R}$
are now manifested as string interactions. 

The fields in the Yang-Mills theory are the low energy degrees of
freedom of open string field theory on D1 branes.  Holography is
realized as the metamorphosis of the fields $X^i$ of the YM theory
into transverse coordinates in ten dimensional space-time. Note that
this space-time interpretation is valid only when $g_s \ll 1$. For
finite $g_s$ the Yang-Mills theory of course makes perfect sense - but
there is no natural space-time interpretation of the nonabelian
degrees of freedom.

In \cite{Craps:2005wd} it was argued that a similar Matrix String
Theory may be written down for $Q \neq 0$. The line of reasoning
which leads to this is similar to the Sen-Seiberg argument \cite{senseiberg}, but
more subtle - as explained in \cite{Craps:2005wd}. The action is a
simple modification of (\ref{mone})
\ben
S = \int d\tau \int_0^{2\pi{\tilde R}}d\sigma {\rm Tr}  \{
\frac{e^{-Q\tau}}{2 g_{YM}^2} F_{\tau \sigma}^2
+ \frac{1}{2} (D_\tau X^i)^2 
- \frac{1}{2} (D_\sigma X^i)^2
+ \frac{g_{YM}^2}{4}e^{Q\tau} [ {X^i}, {X^j} ]^2 \} 
\label{mtwo}
\een
Since this is essentially the action of $J$ D1 branes in the
light cone gauge, $\tau$ is the same as the coordinate $x^+$
in the background. Thus, in the far future in light cone time,
the gauge theory is strongly coupled, while near the singularity
at $x^+ \rightarrow -\infty$ the gauge theory is {\em weakly
coupled}. This means that while the theory has a nice interpretation
as a space-time theory with dynamical gravity in the future,
such an interpretation breaks down at $\tau \rightarrow -\infty$ -
precisely the place where there is a null singularity. Here all
the $J^2$ degrees of freedom are relevant and might "resolve" the singularity.

\subsection{IIB Big Bangs}

The Type IIB version of this background shows a richer structure
\cite{Das:2006dr}. The background is once again given by
(\ref{flatmetric}) and (\ref{nulldilaton}) where both $x^-$ and $x^8$
are compact,
\ben
x^- \sim x^- + 2\pi R, \qquad x^8 \sim x^8 + 2 \pi R_B
\label{mfour}
\een
The usual DLCQ Matrix Theory logic then implies that string theory in the sector
with $p_- = J/R$ and $p_8 = 0$ is described by a $SU(J)$ 2+1
dimensional Yang-Mills theory of $J$ D2 branes \cite{Banks:1996my},
\cite{Motl:1997th}. This gauge theory lives on a $T^2$ with sides
\ben
R_\rho = g_B\frac{l_B^2}{R}~~~~~~
R_\sigma = \frac{l_B^2}{R}
\label{mfive}
\een
where $g_B, l_B$ are the string coupling and the string length of the
original IIB theory. The dimensional coupling constant of the
Yang-Mills is
\ben
G_{YM}^2 = \frac{R}{R_\sigma R_\rho} = \frac{RR_B^2}{g_Bl_B^4}
\label{msix}
\een
We will call this theory "Matrix Membrane Theory".

The action of this matrix membrane theory is given by
\ben
S = \int d\tau \int_0^{2\pi R_\rho} d\sigma
\int_0^{2\pi R_\sigma} d\rho~{\cal L}
\label{six}
\een
where
\bea
{\cal L} = {\rm Tr}  \{  \frac{1}{2}[(D_\tau X^a)^2
- (D_\sigma X^a)^2 - e^{2Q\tau} (D_\rho X^a)^2] 
& + & \frac{1}{2(G_{YM}e^{Q\tau})^2}[F_{\sigma\tau}^2
+ e^{2Q\tau}(F_{\rho\tau}^2- F_{\rho\sigma}^2)]  \nn \\
& + &  \frac{(G_{YM}e^{Q\tau})^2}{4} [ X^a , X^b ]^2 \},
\label{mmembraneaction}
\eea
where $X^a, a= 1 \cdots 7$ are now seven scalar fields
and $F_{\mu\nu}$ denotes the gauge field strength.
Note that there is a factor of $e^{Q\tau}$ with each
$\partial_\rho$ or a covariant vector component $V_\rho$,
in addition to a factor of $e^{Q\tau}$ for each $G_{YM}$

For $Q=0$ and $g_B \ll 1$ this action reproduces the worldsheet action
for Type IIB strings in the light cone gauge. In this limit the
commutator terms force the fields to be diagonal.  The gauge field
strengths can be dualized to a scalar which we will call $X^8$, so
that we have a $2+1$ dimensional action of eight scalar fields. Finally, since
for small $g_B$ we have $R_\rho \ll R_\sigma$, the action reduces to a
1+1 dimensional action which may be then identified with the
Green-Schwarz light cone worldsheet theory. Once again sectors of
boundary conditions describe upto $J$ strings with the spatial extent
of the worldsheeet proportional to their longitudinal momenta.

This story changes interestingly when $Q \neq 0$. The mass scale
associated with the Kaluza Klein modes in the $\rho$ direction is
given by $M_{KK} \sim \frac{R}{g_B l_B^2}$ while the mass scale which
determines the non-abelian dynamics is $G_{YM}$ given in
(\ref{msix}). Thus for $R_B \gg l_B$ the KK modes are much lighter
than the Yang-Mills scale.  In our present time-dependent context,
these scales become time-dependent and it follows from the coupling
and the $\partial_\rho$ terms in (\ref{mmembraneaction}) that the KK
modes are expected to decouple much {\em later} than the time when the
non-abelian excitations decouple. Therefore, there is a regime where
we can ignore the non-abelian excitations, but cannot ignore the KK
modes. In this regime, the Matrix Membrane lagrangian density is given
by 
\ben {\cal L}_{diag} = \frac{1}{2}[\sum_{I=1}^8 (\partial_\tau
X^I)^2 -(\partial_\sigma X^I)^2 - e^{2Q\tau} (\partial_\rho X^I)^2] -
2\mu^2\sum_{I=1}^8 (X^I)^2
\label{ten}
\een
It is tempting to argue that as $\tau \rightarrow
\infty$ the Kaluza-Klein modes in the $\rho$ direction become
infinitely massive, so that the theory becomes $1+1$ dimensional 
and exactly identical to the Green-Schwarz string action in this 
background. However, this is too hasty since we have a time-dependent
background here and energetic arguments do not apply. 

Instead, we
should ask whether {\em any} state at an early time evolves into
a state of the perturbative fundamental string - i.e. states which
do not carry any momentum in the $\rho$ direction. The modes
of the field $Y^I(\rho,\sigma,\tau)$ which are positive frequency 
at early times are given by
\ben
\varphi^{({in})}_{m,n} = 
\{ {\frac{R}{8 \pi^2 l_B^4 g_B}} \}^{1/2}~
\Gamma (1-i\omega_m / Q)~e^{i(\frac{mR}{l_B^2}\sigma
+\frac{n R}{g_B l_B^2}\rho)}~J_{-i\frac{\omega_m}{Q}}
(\kappa_n e^{Q\tau})
\label{inmodes}
\een
where
\ben
\omega_m^2 = \frac{m^2R^2}{l_B^4}~~~~~~~~~~~
\kappa_n = \frac{nR}{Qg_B l_B^2}
\een
while those which are appropriate at late times are
\ben
\varphi^{({out})}_{m,n} = 
\{ \frac{R}{16 \pi l_B^4 g_B Q}\}^{1/2}~
e^{i(\frac{mR}{l_B^2}\sigma
+\frac{n R}{g_B l_B^2}\rho)}~H^{(2)}_{-i\frac{\omega_m}{Q}}
(\kappa_n e^{Q\tau})
\label{outmodes}
\een
The problem at hand is identical to that of a bunch of
two dimensional scalar field (living on $\tau,\sigma$
spacetime) with time dependent masses. It is well
known that such time dependence leads to particle
production or depletion \cite{Strominger:2002pc},
\cite{Birrell:1982ix},\cite{Tanaka:1996cz}. 
Because of standard relations between the Hankel function $H^{(2)}_\nu(z)$
and the Bessel function $J_{\nu}(z)$ there is a non-trivial Bogoliubov
transformation between these modes which imply that the vacua defined
by the in and out modes are not equivalent. 
In fact, the out vacuum $|0>_{out}$ is a squeezed state of the
"in" particles. In other words, {\em if
we require that the final state at late times does not contain any of
the KK modes, the initial state must be a squeezed state of these
modes}. The occupation number of the in modes in the out state is thermal
\ben
_{out}<0|~a^{\dagger I,({in})}_{m,n} a^{I,({in})}_{m
,n} |{0}>_{out} = \frac{1}{e^{\frac{2\pi\omega_m}{Q}}-1}
\een
Note that the Bogoliubov
coefficients and number densitites depend only on $m$ for all
$n \neq 0$. This follows from the fact that $n$- dependence may
be removed by shifting the time $\tau$ by $\log (\kappa_n)$.
However, the modes with $n=0$ need special treatment.  Indeed, in the $n
\rightarrow 0$ limit the ``in'' modes (\ref{inmodes}) go over to
standard positive frequency modes of the form $e^{-i\omega_m \tau}$ as
expected. In this limit, however, the out modes (\ref{outmodes})
contain both positive and negative frequencies. This is of course a
wrong choice, since for these $n=0$ modes there is no difference
between ``in'' and ``out'' states.  In fact, the ``out'' modes
(\ref{outmodes}) have been chosen by considering an appropriate large
time property for {\em nonzero n} and do not apply for $n=0$.
In other words, the squeezed state contains only
the $n \neq 0$ modes.

The operators $a^I_{m,n}$ in fact create states of $(p.q)$ strings 
in the original Type IIB theory
\cite{Banks:1996my}. To see this, let us recall how the light cone IIB 
fundamental string states arise from the $n=0$ modes of the Matrix
Membrane. In this sector, the action is exactly the Green-Schwarz
action. The oscillators $a^{\dagger I}_{m,0}$ defined above are
in fact the world sheet oscillators and create excited states of a
string. The gauge invariance of the theory allows nontrivial boundary
conditions, so that $m$ defined above can be fractional. Equivalently
the boundary conditions are characterized by conjugacy classes of the
gauge group. The longest cycle corresponds to a single string whose
$\sigma$ coordinate has an extent of $2\pi J \frac{l_B^2}{R}$ which is
the same as $2\pi l_B^2 p_-$ as it should be in the light cone gauge.
Shorter cycles lead to multiple strings - the sum of the lengths of
the strings is always $2\pi l_B^2 p_-$, so that there could be at most
$J$ strings. 
Note that $m$ is the momentum in the $\sigma$ direction :
a state with {\em net} momentum in
the $\sigma$ direction in fact corresponds to a fundamental IIB string
wound in the $x^-$ direction. This may be easily seen from the chain
of dualities which led to the Matrix Membrane. 

As shown in \cite{Banks:1996my}, following the arguments of
\cite{Schwarz:1995dk,sM,Aspinwall:1995fw}, $SL(2,Z)$ transformations on
the torus on which the Yang-Mills theory lives become the $SL(2,Z)$
transformations which relate $(p,q)$ strings in the original IIB theory.
In particular, the oscillators $a^I_{0,n}$ create states of a 
D-string.

The state $|0>_{\rm out}$ therefore contain excited states of these
$(p,q)$ strings. The number of such strings depends on the choice of
the conjugacy classes characterzing boundary conditions.  Since each
$(m,n)$ quantum number is accompanied by a partner with $(-m,-n)$ this
state does not carry any F-string or D-string winding number. Finally
this squeezed state contains only $n \neq 0$ modes, i.e. they do not
contain the states of a pure F-string.
We therefore conclude that in this toy model, the initial state has to
be chosen as a special squeezed state of {\em unwound} $(p,q)$ strings
near the big bang to ensure that the late time spectrum contains only
perturbative strings.

\subsection{pp-wave Big Bangs}

The nonabelian degrees of freedom of Matrix String Theory or
Matrix Membrane theory become important near the ``singularity''.
In the background considered above, this theory has one length scale - 
given by the Yang-Mills coupling $G_{YM}^{-1}$. It would be worthwhile 
to find similar situations with an additional length scale with the hope that
tuning the dimensionless ratio would allow us to go to a regime where
some class of nonableian configurations become important. One such example 
is provided by pp-waves \cite{Das:2006dr},\cite{Das:2005vd}.
The string frame metric (\ref{flatmetric}) is now modified to \footnote{The coordinates used here make a space-like isometry explicit \cite{Michelson:2002wa}}
\ben
ds^2  =  2dx^+dx^- - 4\mu^2[(x^1)^2+\cdots(x^6)^2](dx^+)^2 
- 8\mu x^7 dx^8 dx^+ + [(dx^1)^2 + \cdots (dx^8)^2]
\een
The dilaton remains the same as (\ref{nulldilaton}), and there is
an additional 5-form field strength 
\ben
F_{+1234}  =  F_{+5678} = \mu~e^{Qx^+} 
\een
For $Q=0$, the matrix membrane theory has been considered in 
\cite{Gopakumar:2002dq}. The detailed
action in this background has been
derived in \cite{Michelson:2004fh} and \cite{SheikhJabbari:2004ik}
The matrix membrane action for $Q \neq 0$ 
now has additional terms \cite{Das:2006dr}
\bea
{\cal L} = \Tr & \{ & \frac{1}{2}[(D_\tau X^a)^2
- (D_\sigma X^a)^2 - e^{2Q\tau} (D_\rho X^a)^2]
+ \frac{1}{2(G_{YM}e^{Q\tau})^2}[F_{\sigma\tau}^2
+ e^{2Q\tau}(F_{\rho\tau}^2- F_{\rho\sigma}^2)] \nn \\
& - & 2\mu^2[(X^1)^2 + \cdots (X^6)^2 + 4 (X^7)^2] 
+\frac{(G_{YM}e^{Q\tau})^2}{4} [ X^a , X^b ]^2 \nn \\
& - & \frac{4\mu}{(G_{YM}e^{Q\tau})}e^{Q\tau} X^7~F_{\rho\sigma}
- 8\mu i (G_{YM}e^{Q\tau})X^7~[ X^5 , X^6 ]~ \},
\label{mmembraneaction2}
\eea
The new length scale is now $\mu$.

Let us briefly recall the physics of this model for $Q = 0$.
When the original IIB theory is weakly coupled, 
$g_B \ll 1$ with $\frac{\mu l_B^4}{R R_B^2} \sim O(1)$,
the effective coupling constant of this
YM theory is strong. Then, along the lines of the discussion
in the previous subsection, the action becomes identical to the 
worldsheet action for Green-Schwarz string in the pp-wave background
\footnote{The dualization required to convert the gauge field to a scalar 
involves a time dependent rotation \cite{Michelson:2004fh}.}. In fact,
as shown in \cite{Gopakumar:2002dq}, integrating out the Kaluza Klein 
modes in the $\rho$ direction generates string couplings with exactly
the correct strength.
 
It is straightforward to see that one could rescale the fields
and the coordinates to write the lagrangian ${\cal L}$ in the form
\ben
{\mathcal L} = \frac{\mu}{G_{YM}^2} ~{\mathcal L}~(\mu=1, G_{YM}=1)
\een
Therefore, in the limit $\lambda \gg 1$ the Yang-Mills theory becomes
weakly coupled and nonabelian classical solutions play a significant role.
These classical solutions are {\em fuzzy ellipsoids} discussed
in \cite{Michelson:2004fh,Michelson:2005iib} similar to fuzzy spheres 
in M theory and Type IIA pp-waves \cite{fuzzy},
\bea
X^5 &= 2 \sqrt{2} \frac{\mu l_p^3}{R} J^1, & \nn \\
X^6 &= 2 \sqrt{2} \frac{\mu l_p^3}{R} J^2, & \nn \\
X^7 &= 2 \frac{\mu l_p^3}{R} J^3,
\eea
where $J^a$ obey the SU(2) algebra, and the remaining matrices $X^i$ vanish.
These solutions have vanishing light cone energy and 
can be shown~\cite{Michelson:2004fh,Michelson:2005iib}
to preserve all 24 supercharges of the M-theory background.
A detailed study of all the 1/2 BPS states of this model appear in

In the original Type IIB description they are fuzzy D3 branes with a topology
$S^2 \times S^1$ where the $S^1$ factor is the compact space direction.

For $Q \neq 0$ the coupling is always weak near $\tau \rightarrow -\infty$ 
so that these fuzzy ellipsoids proliferate. As $\tau$ increases the
coupling gets stronger and one would expect that they should not be
present, leaving behind only perturbative abelian degrees of freedom
representing the fundamental string. This indeed happens. The size of the
ellipsoids is now time dependent : with some initial size the equations
of motion may be used to examine the size at later times. Numerical
results \cite{Das:2006dr} show that with generic initial conditions,
the size oscillates with an amplitude decaying fast with time. In other
words, at late times we are left with only the abelian configurations which 
can be now interpreted as fundamental strings. The phenomenon of
production/depletion of $(p,q)$ strings is identical to the
$\mu =0$ case described in the previous subsection.

\subsection{Issues}

The key feature of holographic models of this type is that 
conventional space-time is an {\em emergent} phenomenon in a very
special regime. In matrix theories, this is the regime where the
gauge theory coupling is strong so that the fields of the theory
can be interpreted as space-time coordinates of a point on 
a fundamental string. In the toy models of cosmology described
above, such an interpretation appears to be valid at late times.
If we forcibly extrapolate this interpretation to early times 
we encounter a singularity. At this singularity, however, the
holographic gauge theory is {\em weakly} coupled : as such a
space-time interpretation is not valid in any case. Since the
coupling is weak there is a good chance that we have a 
well defined time evolution.

There are several caveats in this general story. The success of Matrix
Theory generally depends on supersymmetry. Even though the backgrounds
considered have half of the supersymmetries, the matrix theory does
not. One of the consequences of this is that a potential for the
fields $X^a$ could be generated which spoils the interpretation in
terms of space-time coordinates. This issue has been investigated in
\cite{Craps:2006xq} and \cite{Li:2005ai}. Indeed there is a potential
at one loop. However it turns out that at late times the potential
vanishes fast, indicating that $X^a$ become moduli \footnote{In
\cite{Li:2005ai} it is claimed that the potential in fact
vanishes. However it turns out that the quantity which is computed in
this paper is a time averaged potential rather than the time dependent
potential \cite{Craps:2006yb}}.

An important question relates to backreaction. Sometimes null
singularities of the type described here are unstable under
perturbations. In the past, orbifold singularities of this type have
been investigated as possibly consistent backgrounds for {\em
perturbative string theory}. However, it was soon found that these
null singularities turn spacelike under small perturbations - large
curvatures develop invalidating the use of perturbative string theory
\cite{Horowitz:2002mw}. In our case, the significance of such an
instability, if present, is rather different. Here the string theory
is in any case strongly coupled near the singularity and there is no
question of a perturbative description. Rather the correct description
is provided by a weakly coupled Yang-Mills theory. The question now is
to find out the meaning of a bulk instability in the gauge theory.  It
remains to be seen if this causes any problem even though the coupling
is weak.  This issue is particularly significant for variations of
this model based on null branes \cite{Robbins:2005ua}.
For other discussions of Matrix Theory in such backgrounds,
see \cite{Ishino:2005ru}.

Perhaps the most important question is about continuation through the
singularity. Even though the holographic theory is weakly coupled near
the null singularity, the hamiltonian expressed in terms of the
conjugate momenta have a singular behavior as one approaches this
region - and it is not clear whether there is an unambiguous
prescription to continue back in time beyond this point. Recently
\cite{Craps:2007iu} has put forward an interesting proposal to
address this issue.

\section{Null Singularities in the AdS/CFT correspondence}

In many respects the AdS/CFT correspondence is a more controlled
example of the holographic principle. In its simplest setting, 
the correspondence implies IIB
string theory on $AdS_5 \times S^5$ with a constant 5-form flux
is dual to $3+1$ dimensional $N=4$ supersymmetric $SU(N)$ Yang-Mills theory
which lives on the boundary of $AdS_5$. If $R_{AdS}$ denotes the 
radius of the $S^5$ as well as the curvature length scale of $AdS_5$
and $g_s$ denotes the string coupling, the coupling constant $g_{YM}$
and the rank of the gauge group $N$ of the Yang Mills theory are 
related by
\ben
\frac{R_{AdS}^4}{l_s^4} = 4\pi g_{YM}^2 N~~~~~~~~~~~g_s = g_{YM}^2
\een
This immediately implies that the gauge theory
describes {\em classical} string theory in the 't Hooft limit
\ben
N \rightarrow \infty~~~~~~g_{YM} \rightarrow 0
~~~~~~~~~~g_{YM}^2N = {\rm finite}
\een

The low energy limit of the closed string theory - supergravity - is a
good approximation only in the strong coupling regime $g_{YM}^2 N \gg
1$.  For small $g_{YM}^2 N$ supergravity and hence conventional
space-time is not a good description of the gauge theory
dynamics. Finite $N$ corrections correspond to string loop effects.

There have been several approaches to cosmological singularities
by finding appropriate modifications of the $AdS$ solutions which
correspond to deformations of the Yang-Mills theory or to states in
the theory \cite{others1}. We will discuss
one approach developed in \cite{Das:2006pw,Das:2006dz,Chu:2006pa,
Lin:2006ie,Lin:2006sx}
\footnote{See \cite{Kraus:2002iv} for an interesting approach to find
signatures of space-like singularities inside AdS black holes in the 
CFT}. 
The hope is similar to that in
the Matrix Theory approach. The idea is to find bulk solutions which
have cosmological singularities where the usual notions of space-time
break down, while the gauge theory description remains tractable.

In the following we will recount the main points in \cite{Das:2006pw,Das:2006dz}. 

\subsection{The Supergravity Background and the Conjecture} 

The usual $AdS_5 \times S^5$ solution is given by the Einstein frame
metric in Poincare coordinates
\ben
ds^2=(\frac{r^2}{R_{AdS}^2}){\eta}_{\mu\nu} dx^\mu dx^\nu +
(\frac{R_{AdS}^2}{r^2})dr^2 + R_{AdS}^2 d\Omega_5^2
\een
and a 5-form field strength and dilaton $\Phi$
\ben
F_{(5)}=R_{AdS}^4(\omega_5 + *_{10} \omega_5)~~~~~\Phi = {\rm constant}
\een
This has maximal supersymmetry.

We consider supergravity solutions which are {\em non-normalizable}
deformations of this,
\bea
\label{mseven}
ds^2 & = & (\frac{r^2}{R_{AdS}^2}) {\tilde g}_{\mu\nu}(x) dx^\mu dx^\nu + 
(\frac{R_{AdS}^2}{r^2})dr^2 + R_{AdS}^2 d\Omega_5^2\ , \nn \\
\Phi & = & \Phi (x) \nn \\
F_{(5)}& = & R_{AdS}^4(\omega_5 + *_{10} \omega_5)
\eea
The equations of motion then imply that the Ricci tensor ${\tilde
R}_{\mu\nu}$ constructed from the metric ${\tilde g}_{\mu\nu}(x)$ must
obey the equation
\ben
{\tilde R}_{\mu\nu} = \frac{1}{2} \partial_\mu\phi \partial_\nu\Phi, 
\label{meight}
\een
while the dilaton must satisfy
\ben
\partial_\mu (\sqrt{-\det({\tilde g})}\ {\tilde g}^{\mu \nu} 
\partial_\nu \Phi) = 0.
\label{mnine} 
\een
It turns out that this solution is the near-horizon limit of the
geometry produced by three branes whose worldvolume metric is given by
${\tilde g}_{\mu\nu}(x)$.

For generic ${\tilde g}_{\mu\nu}(x)$ this solution will have curvature
singularities at the Poincare horizon at $r = 0$. This does not happen
when ${\tilde g}_{\mu\nu}(x)$ and $\Phi (x)$ are functions of a null
coordinate $x^+$. We will therefore restrict our attention to such
solutions. Such solutions retain half of the supersymmetries with
parameters $\epsilon$ satisfying $\Gamma^+ \epsilon = 0$. Furthermore,
for reasons which will become clear soon, we will consider brane
metrics which are conformal to flat space
\ben
{\tilde g}_{\mu\nu}(x) dx^\mu dx^\nu = e^{f(x^+)}~\left[ -2dx^+dx^-
+dx_1^2 + dx_2^2 \right]
\label{mten}
\een
The equations of motion (\ref{meight}) then require that the
dilaton is also a function of $x^+$ alone. The dilaton equation
(\ref{mnine}) is automatically satisfied, while (\ref{meight})
simplifies to
\ben
\label{condnull}
\frac{1}{2}(f')^2 - f'' = \frac{1}{2}(\del_+ \Phi)^2.
\een
where prime denotes derivative with respect to $x^+$.

The conjecture is that string theory in this null background is 
dual to $3+1$ dimensional $N=4$ Yang-Mills theory which lives
on a background space-time given by ${\tilde g}_{\mu\nu}(x)$
and a $x^+$ dependent coupling 
\ben
g_{YM} (x^+) = e^{\Phi(x^+)/2}{\sqrt{g_s}}
\label{meleven}
\een
The bosonic part of the action is
\ben
S = \int d^4 x ~{\rm Tr} \{ \frac{1}{4} 
e^{-\Phi} F_{\mu\nu}F^{\mu\nu}
+ \frac{1}{2} (D_\mu \chi^I)(D^\mu \chi^I) + \frac{1}{4} [ \chi^I , \chi^J
]^2  \}
\label{mtwelve}
\een
where $\chi^I, I = 1,\cdots 6$ are the adjoint scalars.

There are several pieces of evidence for the validity of this
conjecture. First, when $f \ll 1$, we also have $\Phi \ll 1$. In that
case, the solution represents small non-normalizable metric and dilaton
deformations of
standard $AdS_5 \times S^5$.  AdS/CFT correspondence then
implies that the dual gauge theory is deformed by operators which are
dual to these modes, viz. the energy momentum tensor $T_{\mu\nu}$ and
${\rm Tr}F^2$ respectively. This is evident from the action
(\ref{mtwelve}). Secondly, we may consider the action of a single
probe D3 brane in the background and examine the way $f(x^+)$ and
$\Phi(x^+)$ sppear - it is easy to check that this is consistent with
(\ref{mtwelve}). Finally, as noted above, our solution is the
near-horizon limit of the asymptotically flat geometry of a stack of
D3 branes with a curved brane wiorldvolume \footnote{ The full
supergravity solution is given in \cite{Das:2006pw}.}.

We will consider solutions which are $AdS_5 \times S^5$ in asymptotic
null time $x^+ \rightarrow \pm \infty$, but develop null singularities
for some value of $x^+$ which may be chosen to be at $x^+ = 0$. However,
we will require that $g_s \ll 1$ {\em and 
the effective string coupling $e^\Phi g_s$ remains
weak for all $x^+$.}. This latter feature distinguishes our solution
from some others in the literature \footnote{These include orbifold
models, backgrounds with time dependent warping,
models based on tachyon condensation. References to the original
literature can be found in \cite{Das:2006pw}. Some of these topics
are reviewed in \cite{Berkooz:2007fe}.}. A nice example of such a solution is 
\ben
e^{f(x^+)} = \tanh^2x^+~~~~~~~~~~
e^{\Phi}=g_s \left|\tanh \frac{x^+}{2}\right|^{\sqrt{8}}.
\label{mfourteen}
\een
At $x^+ = 0$ all local curvature invariants are bounded. However this 
point may be reached in a finite physical time. For example the affine
parameter $\lambda$ along a geodesic $x^+ (\lambda)$ with all other
coordinates constant is given by
\ben
\lambda = x^+ - \tanh~x^+
\label{mfifteen}
\een
Thus $x^+=0$ can be reached in a finite affine
parameter. Furthermore, it turns out that tidal forces between neighboring geodesics
diverge at this point. Therefore $x^+=0$ is a genuine null
singularity. 

Consider the solution as a time evolution in light
cone time $x^+$. At $x^+ \rightarrow -\infty$ the Yang-Mills coupling
approaches ${\sqrt{g_s}}$ exponentially. In the dual Yang-Mills theory,
we will always work in the 't Hooft limit $g_s \rightarrow 0, N
\rightarrow \infty$ with $g_s N$ finite and large. Therefore, according
to the usual AdS/CFT correspondence, the ground state of the theory is
dual to {\em supergravity} in  $AdS_5 \times S^5$ as stated
above. This vacuum evolves in time according to the Yang-Mills
hamiltonian whose effective coupling {\em decreases}. The dual
description of this time evolution is the supergravity solution
described above. Supergravity, however, makes sense only when the 
Yang-Mills coupling is large. Thus as we approach $x^+ \rightarrow 0$
the coupling approaches zero and the supergravity description becomes
meaningless. The singularity therefore appears at a place where 
we expect a space-time interpretation of the gauge theory to break
down.

\subsection{Some Properties of the Gauge Theory Dual}

As emphasized above, one of the salient features of our toy model is
that the gauge theory is weakly coupled at the "singularity", pretty much like the Matrix Theory examples given above.
This is reassuring, since one would hope that weakly coupled gauge
theory makes sense and provides the alternative structure which
replaces dynamical bulk space time in this region. However, it is
precisely at this point that the nondynamical spacetime of the gauge
theory shrinks to zero size ! 

Normally this would be a disaster since
a gauge theory on a zero size space-time would be singular even if it
is weakly coupled.
What saves the day is the fact that this particular gauge theory is
Weyl invariant. This is evident at the classical level - the factor
$e^{f(x^+)}$ does not appear in the classical action. If the coupling
was constant the theory would have been conformally invariant (in the
sense of invariance under conformal diffeomorphisms) as well. Here the
$x^+$ dependence of the coupling breaks these conformal symmetries but
retains Weyl invariance at least classically. This means that at the
classical level our gauge theory simply does not see the shrinking
Weyl factor.

Usually Weyl invariance of quantum field theories is broken at the
quantum level by anomalies. Our gauge theory is a special case of
$N=4$ Yang-Mills theory coupled to nondynamical conformal supergravity, where only
the metric and the dilaton fields of the background
supergravity are turned on.  The Weyl anomaly of this theory has been
worked out a while ago with the result
\cite{Fradkin:1983tg,Liu:1998bu,Nojiri:1998dh}. 
\ben
<T^\mu_\mu>  =  -\frac{N^2}{64\pi^2} \{ 2(R_{\mu\nu}R^{\mu\nu}
-\frac{1}{3}R^2) 
 +  4 \left[-2(R^{\mu\nu}+\frac{1}{3}Rg^{\mu\nu})\partial_\mu\Phi
  \partial_\nu \Phi + (\nabla^2 \Phi)^2 
+ \frac{4}{3}(g^{\mu\nu}\partial_\mu\Phi \partial_\nu \Phi)^2
\right] \}
\label{msixteen}
\een
In fact, it turns out that the 
{\em operator} $T^\mu_\mu$ involves 
only scalars made out of conformal supergravity
fields  \footnote{We are grateful to A. Tseytlin for a correspondence
  about this point}. In our case, the
only nonvanishing components of the Riemann are $R_{+i+i}$ with
$i=1,2$ and the only nonvanishing component of $\partial_\mu \Phi$ is
$\partial_+ \Phi$. Since there are no nonvanishing components with a
contravariant $+$ index, we cannot form a scalar by
contracting these tensors. Thereofore the Weyl anomaly vanishes for
our null background. This implies that correlation functions of
dressed conformal operators are equal to those in a {\em flat} metric
with the same $x^+$ dependent coupling, 
\ben 
\langle
\prod_a~e^{f(x_a^+)\Delta_a}~{\cal O}_a(x^a)
\rangle_{e^f\eta_{\mu\nu},\Phi(x^+)} = \langle \prod_a~{\cal O}_a(x^a)
\rangle_{\eta_{\mu\nu},\Phi(x^+)}
\label{mseventeen} 
\een 
where $\Delta_a$ is the
conformal dimension of the operator ${\cal O}_a$. In other words, the
shrinking conformal factor is invisible to these observables at the
quantum level.

We are therefore left with a gauge theory on flat space with a $x^+$
dependent coupling. The coupling, however, appears as a overall factor
in the gauge field term. Generally, this would imply that the
propagator of canonically normalized fields would be unconventional.
This could be a danger since the derivatives of $\Phi (x^+)$ diverge
at $x^+ = 0$. Luckily this does not happen either. To see this, fix a
light cone gauge \cite{lightcone1}
\ben A_- = 0 
\een 
The fields $A_+$ are then
determined in terms of the transverse components by a constraint
equation which turns out to be identical to that for the standard
$N=4$ theory {\em by virtue of the fact that the coupling depends only
on $x^+$}, 
\ben \frac{1}{2} \partial_- A_+ = \partial_i A_i +
\frac{i}{\partial_-} \left[ A_i , \partial_- A_i \right]
\label{meighteen}
\een
Let us now define new fields $\bA_i, \bA_+$ as follows
\ben
\bA_i (x) = e^{-\Phi (x^+)/2} A_i(x)~~~~\bA_+ (x) = e^{-\Phi (x^+)/2}
A_+(x) 
\een
Since $\Phi$ is a function of $x^+$ alone, the equation
(\ref{meighteen}) is identical with the replacement $A_i \rightarrow
\bA_i,~A_+ \rightarrow A_+$. In terms of these new fields it may be
easily checked that upto terms which are quadratic in the fields,
\ben
e^{-\Phi(x^+)} {\rm Tr} F^2 = {\rm Tr} \bF^2 - \frac{1}{2} \partial_-
\left[ (\partial_+ \Phi) \bA^i \bA_i \right]
\een
where $\bF$ is the field strength constructed out of $\bA$.
Since the additional term is a total derivative, it does not
contribute to the action. {\em This means that the quadratic terms in the
action are identical to that in the light cone gauge action for
standard $N=4$ theory}. The factors of $e^\Phi$ and its derivatives
appear only in the interaction terms of the $\bA$ fields.
Since the coupling $e^{\Phi/2}$ approaches zero at the singularity and
is small and bounded everywhere else, one might expect that the
correlation functions of the fields $\bA_\mu$ are well behaved. 

Generically, time dependent backgrounds lead to particle production. An initial vacuum state typically evolves into a squeezed state of particle - antiparticle pairs. In our null background, however, such processes do not occur. The argument relies on the fact that in light front quantization the states are labelled by $k_-, k_1, k_2 $ where
$0 \leq k_- \leq \infty$ and $-\infty \leq k_1, k_2 \leq \infty$. Since the background depends only on $x^+$, the momentum along $x^-$,
$k_-$ is conserved. The fock vacuum of the theory has $k_-=k_i = 0$. It is then clear that this state cannot evolve into a state containing particles with nonzero $k_-$ since there are no states with negative $k_-$. The $k_-=0$ sector, however, may cause problems with this argument.  

\subsection{The worldsheet theory}

Since the effective 't Hooft coupling of the Yang-Mills theory becomes small at the singularity, it is natural to expect that stringy effects are large. At the same time, in the large N limit string loop effects should be small as well. It is of interest to investigate whether worldhseet string theory could make sense in this background. Unfortunately, because of the presence of RR flux, we do not have a tractable worldsheet formulation of the full worldsheet theory. However a look at the bosonic part of the action in a physical gauge makes it clear that stringy effects are important near the singularity. For this purpose, let us write the metric in slightly different coordinates 
\ben
ds^2 = \frac{1}{y^2} \left[ e^{f(x^+)} \{ 2dx^+dx^- +d{\vec x}^2 \} + d{\vec y}^2 \right]
\een
where ${\vec y}=(y^1 \cdots y^6)$. Fixing the light cone gauge $x^+ = \tau$ following \cite{lightcone} the bosonic part of the action becomes
\ben
S = \frac{1}{2} \int d\sigma d\tau
 \left[ (\partial_\tau {\vec{x}})^2 +  e^{-f(\tau)}(\partial_\tau
 {\vec{y}})^2
-\frac{1}{y^4}e^{2f(\tau)}e^{\Phi(\tau)}(\partial_\sigma {\vec{x}})^2
-\frac{1}{y^4}e^{f(\tau)}e^{\Phi(\tau)}(\partial_\sigma {\vec{y}})^2
\right],
\een
Since both $e^f$ and $e^\Phi$ vanish at $\tau = 0$, the spatial gradient terms become small here, which implies that stringy modes are not suppressed. It is not clear as yet whether the worldsheet theory is non-singular.

\subsection{Penrose Limits and Matrix Theory}

To learn a little more about the string theory in the bulk it is useful to consider the Penrose limit of our background. For this purpose it is convenient to rewrite the metric as
\ben
ds^2 = r^2 [-dt^2+dq^2+e^{F(z^+)}(dx_2^2+dx_3^2)]
+ \frac{dr^2}{r^2} +  d\psi^2 + \sin^2 \psi d\Omega_4^2,
\label{bone}
\een
where we have used the affine parameter $z^+$ defined by
$z^+ = \int^{x^+}dx~e^{f(x)}$ along a 
null geodesic instead of $x^+$, and the function $F(z^+)$ is defined by
$F(z^+) = f(x^+(z^+))$.
The coordinates $q,t$ are defined by
\ben
z^+ = \frac{1}{\sqrt{2}} (q+t), \qquad x^- = \frac{1}{\sqrt{2}}(t-q).
\een
Now we zoom on a null geodesic given by
\ben
r  = \sin~U,~~~~~
t = -\cot~U, ~~~~
\psi  = U,
\label{btwo}
\end{equation}
After the usual scaling associated with a Penrose limit and a complicated coordinate transformation the Einstein frame metric
is given by \cite{Das:2006pw} 
\ben
ds^2 = 2 dUdV - [H(U)\vec{X}^2 + \vec{Y}^2](dU)^2
+ d\vec{X}^2+d\vec{Y}^2.
\label{bnine}
\een
In the Penrose limit, the coordinate $U$ is related to the
coordinate $z^+$ by $z^+ - \frac{1}{\sqrt{2}}\cot~U$ and
the function $H(U)$ is determined in terms of
$F(z^+)$ by
\ben
H(U) = 1 -
\frac{[1+2(z^+)^2]^2}{4} \left[ 
\frac{d^2F}{(dz^+)^2}+\frac{1}{2}\bigl(\frac{dF}{dz^+}\bigr)^2 \right]
= 1+\frac{[1+2(z^+)^2]^2}{8}\left( \frac{d\Phi}{dz^+} \right)^2,
\label{beleven}
\een
In terms of these coordinates, the singularity appears at $U=Pi/2$.
Near this point,
\ben 
\label{singnearsing}
H(U) \sim \frac{1}{(U-\frac{\pi}{2})^2}, \qquad
e^{\Phi (U)} \sim (U-\frac{\pi}{2})^{\frac{\sqrt{8}}{3}}.
\een
Thus the Penrose limit of our original space-time is singular as well.
In fact, it turns out that the pp-wave is singular if and only if
the pre-Penrose limit original spacetime is singular \cite{bbop}.

The pp-wave space-time has space-like and null isometries. In a way similar to the null dilaton cosmologies in the previous section, one may write down a matrix membrane theory for such a background which has a compact null direction $x^- \sim x^- + 2\pi R$ and
$x^8 \sim x^8 + 2\pi R_B$. The resulting $2+1$ dimensional Yang-Mills
action is
\bea
{\mathcal L} = {\rm Tr} \frac{1}{2} \{ [(D_\tau \chi^\alpha)^2
& - & e^{\Phi(\tau)} (D_\sigma \chi^\alpha)^2 
- e^{-\Phi(\tau)} (D_\rho \chi^\alpha)^2] \nn \\
& + & \frac{1}{G_{YM}^2}[e^{\Phi (\tau)} F_{\sigma\tau}^2
+ e^{-\Phi(\tau)}F_{\rho\tau}^2- F_{\rho\sigma}^2] \nn \\
 & - & H(\tau)[(\chi^1)^2 + (\chi^2)^2] - (\chi^3)^2 
\cdots (\chi^6)^2 - 4 (\chi^7)^2 \nn \\ 
& + & \frac{G_{YM}^2}{2} [ \chi^\alpha , \chi^\beta ]^2 
+ 2 i G_{YM} \chi^7 [ \chi^5, \chi^6 ]
    + \frac{4}{G_{YM}} \chi^7 F_{\sigma\rho}  \},
\label{mmembraneaction5}
\eea
Unlike the matrix membrane in the linear null dilaton discussed in the 
previous section (i) the Yang-Mills coupling of this model is independent of $\tau$, (ii) both $\partial_\rho$ and $\partial_\sigma$ have time-dependent factors. In the IR, the fields in the theory become commuting and may be chosen to be diagonal and the extent of the $\rho$ direction shrinks to zero size. The lagrangian then reduces to the light cone gauge Green-Schwarz worldsheet lagrangian for the fundamental string in the relevant pp-wave background. An analysis similar to that in section 2.1 now shows that excited modes of both D-strings and fundamental strings are now produced by the time dependent background.

It would be interesting to analyze worldsheet string theory in this time dependent pp-wave. Backgrounds with similar singularities in the {\em string frame} metric have been studied earlier \cite{Papadopoulos:2002bg, bbop}
  . In that case the worldsheet equations of motion are solvable in terms of special functions and certain statements about the validity of string theory could be made. In our background the worldsheet action is quadratic in the fields, but the equations of motion are not readily solvable. Our analysis of the Matrix Theory seems to indicate that nonperturbative physics becomes important.
Nevertheless some insight from the worldsheet theory could be valuable.

\subsection{Issues}

The toy model of null cosmology described in this section might provide an interesting way to resolve a null singularity. In the asymptotic past (in light cone time) the gauge theory has a valid space-time interpretation in terms of supergravity. As we approach $x^+ = 0$, the 't Hooft coupling approaches zero, and the space-time description breaks down. Our investigations {\em suggest} that the weakly coupled gauge theory remains controlled and it is this description which should be used to approach and even continue past the singularity.
Our analysis is not detailed enough to decide whether such a continuation is indeed possible.

Some of our conclusions were based on a treatment in the light cone gauge and in light front quantization. Sometimes light front quantization leads to subtleties with the zero longitudinal momentum mode. These subtleties could give rise to infra-red effects which have to be interpreted suitably. 

As in the Matrix Theory models described in the previous section, backreaction due to an initially smooth perturbation which corresponds to a nice {\em normalizable} initial wavefunction of the gauge theory is an interesting question. Note that unlike the Matrix theory example, the bulk string coupling is small near the singularity, though stringy effects are large.
If such a perturbation results in a curvature singularity in the bulk, perturbative string theory will certainly break down here. However, the main idea here is use the gauge theory description in this region.
It remains to be seen whether this causes any problem for the gauge theory inspite of being weakly coupled.
h
These and other questions are currently under investigation.

\section{Acknowledgements}

I would like to thank Jeremy Michelson, K. Narayan and Sandip Trivedi for a very enjoyable collaboration, Adel Awad, Costas Bachas, Ben Craps, David Gross, A. Harindranath, David Kutasov, Gautam Mandal, Shiraz Minwalla, Tristan McLoughlin and Alfred Shapere for discussions at various stages of the research presented here. I am thankful to Amit Sever and Stephen Shenker for correspondence which clarified several issues and Arkady Tseytlin for a correspondence about conformal anomalies. I thank the British University in Egypt in Cairo and Virginia Tech for hosting stimulating meetings. This manuscript was prepared during visits to Tata Institute of Fundamental Research, Mumbai and Indian Association for Cultivation of Science, Kolkata - I thank them for hospitality. The reserach reported here was supported in part by the United States National Science Foundation Grant Numbers PHY-0244811 and PHY-0555444 and Department of Energy contract No.
DE-FG02-00ER45832.


\begin{thebibliography}{99}  

\bibitem{Das:2005vd}
  S.~R.~Das and J.~Michelson,
  Phys.\ Rev.\  D {\bf 72}, 086005 (2005)
  [arXiv:hep-th/0508068].

\bibitem{Das:2006dr}
  S.~R.~Das and J.~Michelson,
  Phys.\ Rev.\  D {\bf 73}, 126006 (2006)
  [arXiv:hep-th/0602099].

\bibitem{Das:2006pw}
  S.~R.~Das, J.~Michelson, K.~Narayan and S.~P.~Trivedi,
  Phys.\ Rev.\  D {\bf 75}, 026002 (2007)
  [arXiv:hep-th/0610053].

\bibitem{Das:2006dz}
  S.~R.~Das, J.~Michelson, K.~Narayan and S.~P.~Trivedi,
  Phys.\ Rev.\  D {\bf 74}, 026002 (2006)
  [arXiv:hep-th/0602107].



\bibitem{'tHooft:1993gx}
  G.~'t Hooft,
  arXiv:gr-qc/9310026;
  L.~Susskind,
  J.\ Math.\ Phys.\  {\bf 36}, 6377 (1995)
  [arXiv:hep-th/9409089].

\bibitem{twod} 
For reviews and references see  I.~R.~Klebanov,
  arXiv:hep-th/9108019;
  S.~R.~Das,
  arXiv:hep-th/9211085; 
  E.~J.~Martinec,
  Class.\ Quant.\ Grav.\  {\bf 12}, 941 (1995)
  [arXiv:hep-th/9412074].

\bibitem{Banks:1996vh}
  T.~Banks, W.~Fischler, S.~H.~Shenker and L.~Susskind,
  Phys.\ Rev.\  D {\bf 55}, 5112 (1997)
  [arXiv:hep-th/9610043].

\bibitem{Susskind:1997cw}
  L.~Susskind,
  arXiv:hep-th/9704080.

\bibitem{Banks:1996my}
  T.~Banks and N.~Seiberg,
  Nucl.\ Phys.\  B {\bf 497}, 41 (1997)
  [arXiv:hep-th/9702187].

\bibitem{Motl:1997th}
  L.~Motl,
  arXiv:hep-th/9701025.

\bibitem{Dijkgraaf:1997vv}
  R.~Dijkgraaf, E.~P.~Verlinde and H.~L.~Verlinde,
  Nucl.\ Phys.\  B {\bf 500}, 43 (1997)
  [arXiv:hep-th/9703030].

\bibitem{Maldacena:1997re}
  J.~M.~Maldacena,
  Adv.\ Theor.\ Math.\ Phys.\  {\bf 2}, 231 (1998)
  [Int.\ J.\ Theor.\ Phys.\  {\bf 38}, 1113 (1999)]
  [arXiv:hep-th/9711200];
  E.~Witten,
  Adv.\ Theor.\ Math.\ Phys.\  {\bf 2}, 253 (1998)
  [arXiv:hep-th/9802150];
  S.~S.~Gubser, I.~R.~Klebanov and A.~M.~Polyakov,
  Phys.\ Lett.\  B {\bf 428}, 105 (1998)
  [arXiv:hep-th/9802109].

\bibitem{twodsingular}
  J.~L.~Karczmarek and A.~Strominger,
  JHEP {\bf 0404}, 055 (2004)
  [arXiv:hep-th/0309138];
  J.~L.~Karczmarek and A.~Strominger,
  JHEP {\bf 0405}, 062 (2004)
  [arXiv:hep-th/0403169];
  S.~R.~Das, J.~L.~Davis, F.~Larsen and P.~Mukhopadhyay,
  Phys.\ Rev.\  D {\bf 70}, 044017 (2004)
  [arXiv:hep-th/0403275];
  S.~R.~Das and J.~L.~Karczmarek,
  Phys.\ Rev.\  D {\bf 71}, 086006 (2005)
  [arXiv:hep-th/0412093];
  S.~R.~Das,
  arXiv:hep-th/0503002;
  S.~R.~Das and L.~H.~Santos,
  Phys.\ Rev.\  D {\bf 75}, 126001 (2007)
  [arXiv:hep-th/0702145].


\bibitem{Craps:2005wd}
  B.~Craps, S.~Sethi and E.~P.~Verlinde,
  JHEP {\bf 0510}, 005 (2005)
  [arXiv:hep-th/0506180].

\bibitem{senseiberg}
  A.~Sen,
  Adv.\ Theor.\ Math.\ Phys.\  {\bf 2}, 51 (1998)
  [arXiv:hep-th/9709220];
  N.~Seiberg,
  Phys.\ Rev.\ Lett.\  {\bf 79}, 3577 (1997)
  [arXiv:hep-th/9710009].




\bibitem{Strominger:2002pc}
  A.~Strominger,
  [arXiv:hep-th{0209090}].

\bibitem{Birrell:1982ix}
  N.~D.~Birrell and P.~C.~W.~Davies,
  {\em{Quantum Fields In Curved Space}}
(Cambridge University Press, Cambridge, 1982).

\bibitem{Tanaka:1996cz}
  T.~Tanaka and M.~Sasaki,
  Phys.\ Rev.\ {\bf 55}(1997) 6061;
  [arXiv:gr-qc/9610060].

\bibitem{sM}
J. H. Schwarz,
Phys.\ Lett. {\bf 367}{1996}{97--103} 
[arXiv:hep-th/9510086]

\bibitem{Schwarz:1995dk}
  J.~H.~Schwarz,
  Phys.\ Lett.\ B {\bf 360}, 13 (1995)
  [Erratum-ibid.\ B {\bf 364}, 252 (1995)]
  [arXiv:hep-th/9508143].

\bibitem{Aspinwall:1995fw}
  P.~S.~Aspinwall,
  Nucl.\ Phys.\ Proc.\ Suppl.\  {\bf 46}, 30 (1996)
  [arXiv:hep-th/9508154].


\bibitem{Michelson:2002wa}
  J.~Michelson,
  ``(Twisted) toroidal compactification of pp-waves,''
  Phys.\ Rev.\ D {\bf 66}, 066002 (2002)
  [arXiv:hep-th/{0203140}].



\bibitem{Michelson:2004fh}
  J.~Michelson,
  arXiv:hep-th/0401050.




\bibitem{Gopakumar:2002dq}
  R.~Gopakumar,
  Phys.\ Rev.\ Lett.\  {\bf 89}, 171601 (2002)
  [arXiv:hep-th/0205174].

\bibitem{Michelson:2005iib}
J.~Michelson - {\em unpublished}


\bibitem{SheikhJabbari:2004ik}
  M.~M.~Sheikh-Jabbari,
  JHEP {\bf 0409}, 017 (2004)
  [arXiv:hep-th/0406214].


\bibitem{SheikhJabbari:2005mf}
  M.~M.~Sheikh-Jabbari and M.~Torabian,
  JHEP {\bf 0504}, 001 (2005)
  [arXiv:hep-th/0501001];
M.~Ali-Akbari, M.~M.~Sheikh-Jabbari and M.~Torabian,
  Phys.\ Rev.\  D {\bf 74}, 066005 (2006)
  [arXiv:hep-th/0606117].


\bibitem{fuzzy}
  K.~Dasgupta, M.~M.~Sheikh-Jabbari and M.~Van Raamsdonk,
  JHEP {\bf 0205}, 056 (2002)
  [arXiv:hep-th/0205185];
  S.~R.~Das, J.~Michelson and A.~D.~Shapere,
  Phys.\ Rev.\  D {\bf 70}, 026004 (2004)
  [arXiv:hep-th/0306270].




\bibitem{Craps:2006xq}
  B.~Craps, A.~Rajaraman and S.~Sethi,
  Phys.\ Rev.\  D {\bf 73}, 106005 (2006)
  [arXiv:hep-th/0601062].

\bibitem{Li:2005ai}
  M.~Li and W.~Song,
  JHEP {\bf 0608}, 089 (2006)
  [arXiv:hep-th/0512335].

\bibitem{Craps:2006yb}
  B.~Craps,
  Class.\ Quant.\ Grav.\  {\bf 23}, S849 (2006)
  [arXiv:hep-th/0605199].

\bibitem{Robbins:2005ua}
  D.~Robbins and S.~Sethi,
  JHEP {\bf 0602}, 052 (2006)
  [arXiv:hep-th/0509204];

\bibitem{Horowitz:2002mw}
  G.~T.~Horowitz and J.~Polchinski,
  Phys.\ Rev.\  D {\bf 66}, 103512 (2002)
  [arXiv:hep-th/0206228];
  A.~Lawrence,
  JHEP {\bf 0211}, 019 (2002)
  [arXiv:hep-th/0205288].



  E.~J.~Martinec, D.~Robbins and S.~Sethi,
  JHEP {\bf 0608}, 025 (2006)
  [arXiv:hep-th/0603104].

\bibitem{Craps:2007iu}
  B.~Craps and O.~Evnin,
  arXiv:0706.0824 [hep-th].


\bibitem{Ishino:2005ru}
  T.~Ishino, H.~Kodama and N.~Ohta,
  Phys.\ Lett.\  B {\bf 631}, 68 (2005)
  [arXiv:hep-th/0509173];
  T.~Ishino and N.~Ohta,
  Phys.\ Lett.\  B {\bf 638}, 105 (2006)
  [arXiv:hep-th/0603215].

\bibitem{others1}T.~Hertog and G.~T.~Horowitz,
  JHEP {\bf 07}, 073 (2004)
  arXiv:.
  T.~Hertog and G.~T.~Horowitz,
  ``Holographic description of AdS cosmologies,''
  JHEP {\bf 04}, 005 (2005)
  [arXiv:hep-th/0503071].



\bibitem{Chu:2006pa}
  C.~S.~Chu and P.~M.~Ho,
  JHEP {\bf 0604}, 013 (2006)
  [arXiv:hep-th/0602054].

\bibitem{Lin:2006ie}
  F.~L.~Lin and W.~Y.~Wen,
  JHEP {\bf 0605}, 013 (2006)
  [arXiv:hep-th/0602124].


\bibitem{Lin:2006sx}
  F.~L.~Lin and D.~Tomino,
  JHEP {\bf 0703}, 118 (2007)
  [arXiv:hep-th/0611139].





\bibitem{Kraus:2002iv}
  P.~Kraus, H.~Ooguri and S.~Shenker,
  Phys.\ Rev.\  D {\bf 67}, 124022 (2003)
  [arXiv:hep-th/0212277];
  L.~Fidkowski, V.~Hubeny, M.~Kleban and S.~Shenker,
  JHEP {\bf 0402}, 014 (2004)
  [arXiv:hep-th/0306170].


\bibitem{Berkooz:2007fe}
  M.~Berkooz, Z.~Komargodski and D.~Reichmann,
  arXiv:0706.0610 [hep-th].

\bibitem{Fradkin:1983tg}
  E.~S.~Fradkin and A.~A.~Tseytlin,
  Phys.\ Lett.\  B {\bf 134}, 187 (1984).

\bibitem{Liu:1998bu}
  H.~Liu and A.~A.~Tseytlin,
  Nucl.\ Phys.\  B {\bf 533}, 88 (1998)
  [arXiv:hep-th/9804083].

\bibitem{Nojiri:1998dh}
  S.~Nojiri and S.~D.~Odintsov,
  Phys.\ Lett.\  B {\bf 444}, 92 (1998)
  [arXiv:hep-th/9810008].

\bibitem{lightcone1}
For an introduction, see
  A.~Harindranath,
  arXiv:hep-ph/9612244.


\bibitem{lightcone}
  R.~R.~Metsaev and A.~A.~Tseytlin,
  Phys.\ Rev.\ D {\bf 63}, 046002 (2001)
  [arXiv:hep-th/0007036];
  R.~R.~Metsaev, C.~B.~Thorn and A.~A.~Tseytlin,
  Nucl.\ Phys.\ B {\bf 596}, 151 (2001)
  [arXiv:hep-th/0009171];
  J.~Polchinski and L.~Susskind,
  ``String theory and the size of hadrons,''
  [arXiv:hep-th/{0112204}].

\bibitem{bbop}
M. Blau, M. Borunda, M. O'Loughlin and G. Papadopoulos,
``Penrose Limits and Spacetime Singularities,''
Class.\ Quant.\ Grav.\ {\bf 21\/}, L43 (2004)
[arXiv:hep-th/{0312029}].

\bibitem{Papadopoulos:2002bg}
  G.~Papadopoulos, J.~G.~Russo and A.~A.~Tseytlin,
  Class.\ Quant.\ Grav.\  {\bf 20}, 969 (2003)
  [arXiv:hep-th/0211289].




\end{thebibliography}
\end{document}